# Unveiling the anisotropic fractal magnetic domain structure in bulk crystal of antiskyrmion-host (Fe,Ni,Pd)$_3$P by small-angle neutron scattering

**Kosuke Karube[a]\*, Victor Ukleev[b], Fumitaka Kagawa[ac], Yoshinori Tokura[acd], Yasujiro Taguchi[a]\* and Jonathan S. White[b]\***

[a]RIKEN Center for Emergent Matter Science (CEMS), Wako, 351-0198, Japan

[b]Laboratory for Neutron Scattering and Imaging, Paul Scherrer Institute, Villigen, CH-5232, Switzerland

[c]Department of Applied Physics, University of Tokyo, Bunkyo-ku, 113-8656, Japan

[d]Tokyo College, University of Tokyo, Bunkyo-ku, 113-8656, Japan

Correspondence email: kosuke.karube@riken.jp; y-taguchi@riken.jp; jonathan.white@psi.ch

**Abstract**    Intermetallic Pd-doped (Fe,Ni)$_3$P, that crystalizes in a non-centrosymmetric tetragonal structure with $S_4$ symmetry, has recently been discovered to host magnetic antiskyrmions, antivortex-like topological spin textures. In this material, uniaxial magnetic anisotropy and dipolar interactions play a significant role, giving rise to finely branched magnetic domain patterns near the surface of bulk crystals, as revealed by a previous magnetic force microscopy (MFM) measurement. However, small-angle neutron scattering (SANS) is a more suitable method for characterizing bulk properties and fractal structures at the mesoscopic length scale. In this study, using SANS and MFM, we quantitatively investigate the magnetic domain structure in bulk single crystals of (Fe$_{0.63}$Ni$_{0.30}$Pd$_{0.07}$)$_3$P. The SANS results demonstrate that the magnetic domain structure exhibits anisotropic fractal character on the length scale down to the width of the magnetic domain walls. The fractal features are gradually lost in magnetic fields, and different field dependences are observed at 300 K and 2 K due to a temperature-dependent anisotropy. This study quantifies the fractality of the highly anisotropic magnetic domain structures in an antiskyrmion material, and highlights the versatility of SANS for the study of fractal structures in magnetic systems.

## 1. Introduction

Small-angle neutron scattering (SANS) is a powerful tool for determining magnetic structures in condensed matter physics (Mühlbauer *et al*., 2019). SANS provides unique information for characterizing bulk properties over the mesoscopic length scale from a few nm to several hundred nm. This is a very important length regime in which various types of magnetic structures emerge. Magnetic domains (Malozemoff & Slonczewski, 1979; Hubert & Schäfer, 1998) and magnetic skyrmions (Mühlbauer *et al*., 2009; Yu *et al*., 2010; Nagaosa & Tokura, 2013) are typical examples of such mesoscopic magnetic structures. Skyrmions are vortex-like spin textures characterized by an integer topological charge, and have been extensively studied in the last decade in various magnetic systems as a source of emergent electromagnetic phenomena, and with an anticipation for potential applications in spintronics devices (Tokura & Kanazawa, 2021).

Recently, an antivortex-like new topological spin texture, the antiskyrmion, has attracted much attention, in which the sign of the topological charge is opposite to that of the skyrmion. Antiskyrmions are topologically distinct to conventional (type I and II) bubble domains and skyrmions, and composed of both Bloch and Néel walls with reversed helicities along two orthogonal axes, i.e., with four-fold roto-inversion ($\bar{4}$) symmetry. This unique magnetic texture originates from the anisotropic Dzyaloshinskii-Moriya interaction (DMI) in the non-centrosymmetric crystal structure belonging to $D_{2d}$ or $S_4$ symmetry (Bogdanov & Yablonskii, 1989; Leonov *et al*., 2016). In contrast to a number of skyrmion materials, antiskyrmions have thus far been observed only in thin plates (100−200 nm thickness) of Heusler alloys with $D_{2d}$ symmetry, e.g., $Mn_{1.4}PtSn$ (Nayak *et al*., 2017; Peng *et al*., 2020), and Pd-doped schreibersite (Fe,Ni)$_3$P with $S_4$ symmetry, i.e., $(Fe_{0.63}Ni_{0.30}Pd_{0.07})_3P$ (Karube *et al*., 2021) by using Lorentz transmission electron microscopy (LTEM). It has been identified that antiskyrmion formation is governed by the competition between uniaxial magnetic anisotropy and dipolar interactions in the presence of the anisotropic DMI. Therefore, the antiskyrmions are square shaped and readily transform into elliptically deformed skyrmions to reduce magnetostatic energy upon changing various parameters (Peng *et al*., 2020; Karube *et al*., 2021; Karube *et al*., 2022).

In $(Fe_{0.63}Ni_{0.30}Pd_{0.07})_3P$, again due to the dominance of uniaxial anisotropy and dipolar interactions, helical and antiskyrmion spin textures several hundreds of nm in size change to micrometer-sized magnetic domain structures as the crystal thickness is increased. Sawtooth-like domain walls and finely branched fractal magnetic domain patterns have been observed near the surface of bulk crystals by using magnetic force microscopy (MFM) (Karube *et al*., 2021). Thickness-dependent complex magnetic textures have been reported also for $Mn_{1.4}PtSn$ (Ma *et al*., 2020; Sukhanov *et al*., 2020; Cespedes *et al*., 2021). MFM studies have provided a clue for understanding the magnetic structure in bulk $(Fe_{0.63}Ni_{0.30}Pd_{0.07})_3P$, which is very different from the LTEM observations of the helical and antiskyrmion spin textures in thin plates. However, MFM imaging technique probes stray magnetic fields at the sample surface and does not detect the fine magnetic structure deep inside the bulk crystal. In addition, the detailed temperature and field dependence of the fractal magnetic domains has not been investigated in the previous MFM study (Karube *et al.*, 2021).

SANS is therefore well suited for the study of fractal magnetic domain structures because it allows quantitative characterization of the thickness-averaged bulk magnetic structures at the mesoscopic length scale, covering the length scales of both fractal magnetic domains and magnetic domain walls (Fig. 1). Small-angle X-ray and neutron scattering are widely used in many fields of science to investigate fractal structures in various systems such as polymers, colloids, porous media, etc., since the power-law exponent of the radial intensity profile $I(q) = I_0 q^{-n}$ (Porod slope) directly determines the fractal dimension: mass fractal dimension $D_m = n$ ($1 \leq n \leq 3$) or surface fractal dimension $D_s = 6-n$ ($3 \leq n \leq 4$), which mathematically quantifies the structural complexity in the system (Schmidt, 1995; Schmidt, 1989; Mildner & Hall, 1986; Martin & Hurd, 1987). The mass fractal describes the density of objects in a given volume, while the surface fractal indicates the roughness of boundary surfaces. For example, $n \sim 4$ ($D_s \sim 2$) in accord with the so-called Porod's law (Porod, 1951) corresponds to smooth surfaces, while $n \sim 3$ ($D_s \sim 3$) indicates fractal surfaces finely branched over multiple length scales to fill the three-dimensional space. Kreyssig *et al.* applied this fractal analysis by SANS to highly anisotropic magnetic domain structures in $Nd_2Fe_{14}B$ single crystals (Kreyssig *et al.*, 2009), where the "surfaces" correspond to the magnetic domain walls. While the original fractal analysis (Schmidt, 1995) was derived assuming isotropic scattering, they applied this analysis to specific regions between anisotropic scattering, allowing an assumption of isotropic scattering in the selected regions. Sukhanov *et al.* have reported a similar SANS study and analysis for $Mn_{1.4}PtSn$ single crystals (Sukhanov *et al.*, 2020). Nevertheless, to the best of our knowledge, there have been no other examples of fractal magnetic domain structures in single crystals studied by SANS.

In this study, SANS measurements on a bulk single crystal of $(Fe_{0.63}Ni_{0.30}Pd_{0.07})_3P$ were performed, in conjunction with field-dependent magnetometry and MFM imaging. As summarized in Table 1, $(Fe_{0.63}Ni_{0.30}Pd_{0.07})_3P$ has similar magnetic properties to $Nd_2Fe_{14}B$ and $Mn_{1.4}PtSn$, in terms of easy-axis magnetic anisotropy along the tetragonal [001] direction and large saturation magnetizations (demagnetization energy), which can be the sources of fractal magnetic domain textures. In fact, anisotropic SANS patterns similar to those observed in previous studies (Kreyssig *et al.*, 2009; Sukhanov *et al.*, 2020) were observed, and data analysis based on Porod slopes was applied. According to the analysis, we present evidence for highly anisotropic fractal properties of the magnetic domains in the bulk, and discuss their temperature and magnetic field dependence in detail.

**2. Experimental**

**2.1. Sample preparations**

Bulk single crystals of $(Fe_{0.63}Ni_{0.30}Pd_{0.07})_3P$ were synthesized by a self-flux method as detailed in the previous study (Karube *et al*., 2022). Pure metals and red phosphorous in an off-stoichiometric molar ratio, Fe : Ni: Pd : P = 1.8 : 1.1 : 0.6 : 1.0, were sealed in an evacuated quartz tube. After a preliminary reaction, the sample was cooled slowly from 970°C to 910°C using a Bridgman furnace. The tetragonal crystal structure with $S_4$ symmetry (space group: $I\bar{4}$) was confirmed by powder X-ray diffraction. The chemical composition of the obtained crystals was determined using energy dispersive X-ray analysis. The single crystals were cut into rectangles along the (110), ($\bar{1}$10) and (001) planes after the crystal orientations were checked by X-ray Laue photography.

**2.2. Magnetization measurements**

Magnetization measurements were carried out using a superconducting quantum interference device magnetometer (MPMS3, Quantum Design). To facilitate comparison with SANS data, the magnetic field in the magnetization curve in Fig. 3(h) was calibrated with multiplying by 1.53, the ratio of the demagnetization factor along the [001] axis of the samples used for the magnetization and SANS measurements.

**2.3. MFM measurements**

MFM measurements were performed at room temperature using a commercial scanning probe microscope (MFP-3D, Asylum Research) with an MFM cantilever (MFMR, Nano World) at a lift height of 50 nm. A plate-shaped bulk single crystal with a thickness of 0.27 mm was used (Supporting information Fig. S1). The flat (001) surface was prepared by chemical mechanical polishing with colloidal silica. In order to apply external magnetic fields perpendicular to the sample plate, the sample was placed on a cylindrical Nd-Fe-B permanent magnet with a diameter of 3 mm and a thickness of 0.5−3 mm. The magnitude of the field at the surface of the permanent magnet (0.15–0.44 T) was measured using a Tesla meter. To facilitate comparison with SANS data, the magnetic fields in Fig. 2 and Supporting information Fig. S1 were calibrated with multiplying by 0.850, the ratio of the demagnetization factor along the [001] axis of the samples used for the MFM and SANS measurements.

**2.4. SANS measurements**

SANS measurements were performed using the instrument SANS-I at the Paul Scherrer Institute (PSI), Switzerland. The schematic configuration of the SANS experiment is shown in Fig. 1(b). A bulk single crystal with a thickness of 0.96 mm was installed into a horizontal-field cryomagnet so that both the neutron beam and the field direction were parallel to the [001] axis. The neutron beam with 9 Å wavelength was collimated before the sample, and the scattered neutrons were detected by a two-dimensional detector. We used the following two configurations: (1) a collimator length of 18 m and a detector length of 20 m to cover the low-*q* region ($\geq$ 0.003 Å$^{-1}$), and (2) a collimator length of 8 m and a detector length of 8 m to cover the high-$q$ region ($\leq 0.04$ Å$^{-1}$). For the SANS patterns in Fig. 3, only data collected using the first configuration are presented. For the intensity profiles as a function of $q$ in Figs. 4 and 5, the data obtained from the two configurations were merged. For all SANS data, background signals from the sample and the instrument were subtracted using the data for the fully polarized ferromagnetic state at 0.8 T. The SANS data before background subtraction in the two configurations are presented in Supporting information Fig. S3. The field-dependent SANS measurements at 2 K and 300 K were performed during a field increasing process. For each measurement, rocking scans were performed, i.e., the cryomagnet was rotated together with the sample, around the vertical [$\bar{1}$10] direction (rocking angle $\omega$) and the horizontal [110] direction (rocking angle $\chi$) in the range from -3° to 3° by 0.2° step. Here, the crystal was precisely aligned so that the origin of the rocking scans ($\omega = \chi = 0°$) corresponds to the beam || [001] configuration. The observed full width at half maximum (FWHM) of the rocking curves was ~ 0.5° for both orientations, as shown in Supporting information Fig. S4. All the SANS patterns displayed in this paper were obtained by summing over the SANS measurements taken at each angle of the rocking scans (-3° $\leq \omega \leq$ 3° and -3° $\leq \chi \leq$ 3°).

## 3. Results and discussion

### 3.1. Imaging of fractal magnetic domains by MFM

First, to visualize the fractal magnetic domains in real space, MFM images at room temperature and at several magnetic fields are presented in Fig. 2 with different magnifications. At zero field, complex stripe domains with a main period of about 20 $\mu$m, aligned along the [110] and [$\bar{1}$10] axes, are observed [Fig. 2(a-c)]. Smaller subdomains with opposite magnetization are embedded in the stripe domains, and the sawtooth-shaped domain walls are finely branched. As detailed in Supporting information S2 and Fig. S2, a box-counting analysis (Falconer, 1990; Smith *et al.*, 1996; Han *et al.*, 2002; Lisovskii *et al.*, 2004) of the MFM image reveals that the domain walls exhibit fractal behaviour over the box length scale from 323 nm to 54 nm, with a fractal dimension of $D = 1.29(1)$. The chirality of the sawtooth pattern is reversed along the two orthogonal axes, which is attributed to the reversed helicity of the Bloch walls deep inside the bulk (Karube *et al.*, 2021). The fast Fourier transform (FFT) of Fig. 2(a) is shown in Fig. 2(d). The dots aligned horizontally with large magnitude correspond to the period of the main stripe domain and its higher harmonics. As indicated with the yellow circles, the intensity distribution of the broad cross-shaped pattern is tilted clockwise from the vertical axis, but anticlockwise from the horizontal axis, respectively, reflecting the asymmetric sawtooth domain pattern with $\bar{4}$ symmetry.

In a magnetic field of 0.18 T, the stripe domains are fragmented and transformed into square, letter N- and letter J-shaped domains, as indicated with the red, blue and green circles, respectively [Fig. 2(e)], which are surrounded by a number of smaller self-similar domains with several length scales [Fig. 2(f, g)]. There is also a $\bar{4}$ symmetry in the square domains themselves and between the N- and J- shaped domains, as in the antiskyrmion spin texture. The FFT pattern is blurred but still shows the asymmetric cruciform shape [Fig. 2(h)]. Therefore, while the topology of the domains has changed significantly, the anisotropic fractal feature remains. At a field of 0.25 T, the largest domains become smaller and rounder, and the smallest bubble-like domains exist more sparsely [Fig. 2(i-k)], indicating that both anisotropic and fractal features are suppressed. Correspondingly, the FFT pattern becomes more isotropic [Fig. 2(l)]. Above 0.35 T, the magnetic domain structure disappears and a single domain state is formed (Supporting information Fig. S1). From the box-counting analysis, the fractal dimensions of the MFM images at 0.18 T and 0.25 T were calculated to be 1.23(1) and 1.22(1), respectively.

### 3.2. Characterization of fractal magnetic domains by SANS

Next, we present SANS results obtained at 2 K and 300 K and quantitative analyses of the fractal magnetic domain structure, and discuss its magnetic field dependence. The results are considered in comparison with other materials. Finally, we also discuss an additional broad peak structure observed only at 300 K.

### 3.2.1. SANS patterns and magnetization

Figure 3(a−f) shows SANS patterns at 2 K and 300 K at several magnetic fields after background subtraction. At zero field, anisotropic diffuse scattering patterns with long tails along the [110] and [$\bar{1}$10] directions are observed both at 2 K and 300 K. Note that this anisotropic scattering is present only within the basal plane with a long correlation length along the easy [001] direction as evidenced from the rocking scans (Supporting information Fig. S4). Under external magnetic fields, the anisotropic SANS intensities gradually disappear. The SANS intensities integrated over the region along the [110] and [$\bar{1}$10] directions are plotted against the magnetic field in Fig. 3(g) and compared with magnetization curves in Fig. 3(h). At zero field, the SANS intensity at 2 K is about twice as large as that at 300 K. The SANS intensities at 2 K and 300 K decrease smoothly with increasing the field and disappear around 0.5 T and 0.3 T, respectively. These critical magnetic fields almost coincide with the saturation fields of the magnetization curve. Therefore, the observed SANS intensities are attributed to scattering from the magnetic domain structures.

Notably, the cross-shaped SANS pattern at zero field is well aligned along the vertical and horizontal directions, except for a slight rotation due to sample misalignment (~ 3°), and there is no asymmetric twisting as observed in the FFT of the MFM image. This is because the sawtooth domain pattern reverses for the top surface and bottom surface when viewed from the same direction [Fig. 1(b)] due to the opposite direction of the stray magnetic fields from the two surfaces, as demonstrated in the previous MFM study and micromagnetic simulations (Karube *et al*., 2021), and thus the asymmetric feature from the sawtooth pattern near the surface becomes smeared in the thickness-averaged magnetic structure probed by SANS. Nevertheless, as will be shown in the next section, there is an anisotropic fractal character in the SANS pattern.

### 3.2.2. Fractal analysis at zero field

First, we analyse SANS data at zero field. The radial SANS intensity profiles $I(q)$ at 2 K and 300 K are plotted in Fig. 4 on a log-log scale. Since the SANS pattern is highly anisotropic, $I(q)$ is determined from the azimuthally integrated intensities over narrow regions along the [110] and [$\bar{1}$10] axes ($q \parallel$ [110]) [Fig. 4(a, c)], and the regions along the [100] and [010] axes ($q \parallel$ [100]) [Fig. 4(b, d)], respectively, so that an analysis based on isotropic scattering can be assumed for each direction. At 2 K, $I(q)$ for $q \parallel$ [110] follows power functions, $I(q) = I_0 q^{-n}$, with two different exponents below and above $q_0 = 0.0138$ Å$^{-1}$, where the slope sharply changes [Fig. 4(a)]. The power function fitting yields $n = 3.09(1)$ for $q < q_0$ (red line) and $n = 5.54(2)$ for $q > q_0$ (blue line). This characteristic SANS intensity profile is similar to those of Nd$_2$Fe$_{14}$B (Kreyssig *et al.*, 2009) and Mn$_{1.4}$PtSn (Sukhanov *et al.*, 2020), indicating that Porod analysis similar to the previous studies can be applied. The values of $n$ and $q_0$ for the three different materials at zero field are summarized in Table 2. As discussed in the literature, the power-law behaviour with $n \sim 3.1$ (surface fractal dimension $D_s \sim 2.9$) in the low-$q$ region below $q_0$ shows evidence for a strongly fractal magnetic domain structure. Following convention and assuming a Gaussian distribution, we can estimate the lower limit of length scales over which fractal domain structures exist according to $2\pi/q_0 \sim 46$ nm. In the high-$q$ region above $q_0$, scattering from the magnetic domain walls is detected. The exponent $n \sim 5.5$ obtained above $q_0$ is much larger than 4, indicating a broad profile at the boundary (Schmidt, 1995; Kreyssig *et al.*, 2009), i.e., the magnetization rotates gradually within thick domain walls. On the other hand, $I(q)$ for $q \parallel$ [100] obeys a single power function without a sharp change in the slope [Fig. 4(b)]. The fitting gives $n = 4.13(2)$ (green line), which is close to the Porod's law. Therefore, the fractal feature is present in $q \parallel$ [110] but not in $q \parallel$ [100], indicating that the fractality of the magnetic domain structure is highly anisotropic.

At 300 K, as compared to the 2 K data, $I(q)$ for $q \parallel$ [110] shows a similar power-law behaviour with a sharp change in the slope below and above $q_0 = 0.0101$ Å$^{-1}$, while a weak bump structure is superimposed on the power-law scattering at the higher $q$ region as indicated with the arrow [Fig. 4(c)]. This high-$q$ bump structure can be seen more clearly under magnetic fields [Fig. 5(c)]. Power-law fitting over the $q$ region without the bump structure yields $n = 3.10(2)$ for $q < q_0$ (red line) and $n = 5.4(1)$ for $q > q_0$ (blue line), which are almost the same values as the 2 K result. Similarly, $I(q)$ for $q \parallel$ [100] obeys a single power function at the low $q$ region, while a clear bump structure is observed at the high $q$ region as in the case of $q \parallel$ [110] [Fig. 4(d)]. The power-law fitting gives $n = 4.12(6)$ (green line), which is again very close to the 2 K data. These almost identical exponents to the 2 K data in both directions indicate that a similar anisotropic fractal magnetic domain structure is realized at 300 K, with an additional isotropic structure present on even smaller length scales, which will be discussed in details in the section 3.2.4.

The crossover length $2\pi/q_0$ (~ 46 nm at 2 K and ~ 62 nm at 300 K) corresponds to the domain wall width (Kreyssig *et al*., 2009). Indeed, this value shows a good agreement with the domain wall width of antiskyrmions (~ 52 nm) at room temperature and zero field observed in real-space imaging using differential phase contrast technique by scanning transmission electron microscopy (Karube *et al*., 2021). The ratio of $2\pi/q_0$ at 2 K to that at 300 K is 0.73. This difference is explained with the formula of temperature ($T$)-dependent domain wall width given by $\delta(T) = \pi\sqrt{A(T)/K_u(T)}$, where $A$ is the exchange stiffness and $K_u$ is the uniaxial anisotropy constant. Theoretical scaling relations with the saturation magnetization ($M_s$), i.e., $A(T) \propto [M_s(T)]^{1.8}$ and $K_u(T) \propto [M_s(T)]^3$, yield $\delta(T) \propto [M_s(T)]^{-0.6}$ (Atxitia *et al*., 2010; Moreno *et al*., 2016). Using the saturation magnetizations at 300 K and 2 K, the ratio of the domain wall width at 2 K to that at 300 K is obtained to be 0.78, which is in good agreement with the ratio of $2\pi/q_0$ (0.73) determined by SANS.

As shown in Table 2, in $Nd_2Fe_{14}B$ and $Mn_{1.4}PtSn$, the exponent becomes $n \sim 3.7$ ($D_s \sim 2.3$), i.e., fractality becomes weaker, below the spin reorientation temperature $T_{SR} = 135$ K and 170 K, respectively, where the magnetic moments cant away from the [001] direction (Kreyssig *et al*., 2009; Sukhanov *et al*., 2020). In contrast, in $(Fe_{0.63}Ni_{0.30}Pd_{0.07})_3P$, the exponent $n \sim 3.1$ ($D_s \sim 2.9$) does not change between 2 K and 300 K due to an absence of a temperature-induced spin reorientation in this composition (Karube *et al*., 2022). These results indicate that easy-axis magnetic anisotropy plays a major role in the formation of fractal magnetic domain structures. The cross-shaped anisotropic SANS pattern is attributed to in-plane magnetocrystalline anisotropy below $T_{SR}$ in $Nd_2Fe_{14}B$ (Pastushenkov *et al*., 1997; Kreyssig *et al*., 2009) and to anisotropic DMI in $Mn_{1.4}PtSn$ and $(Fe_{0.63}Ni_{0.30}Pd_{0.07})_3P$, which fix the principal domain walls perpendicular to in-plane specific directions.

While it is difficult to precisely describe the three-dimensional real-space structure of the magnetic domains with a fractal dimension close to 3, the finely branched domain structure on the surface of the sample, as observed by MFM, presumably penetrates into the bulk to some extent and forms numerous dagger-like domains from the surface to the inner regions, as has long been discussed in the literature (Kaczer & Gemperle, 1960; Hubert & Schäfer, 1998).

### 3.2.3. Magnetic field dependence of fractality

Next, the above fractal analysis is extended to data under magnetic fields. The $I(q)$ profiles at various magnetic fields for $q \parallel [110]$ and $q \parallel [100]$ at 2 K and 300 K are plotted in Fig. 5 in a log-log scale. At 2 K, for $q \parallel [110]$, as the magnetic field is increased, the two different slopes of $I(q)$ below and above $q_0$ become closer while the scattering intensity decreases [Fig. 5(a)]. Note that the kink position at $q_0$ is independent of the field as indicated with the vertical dashed line. On the other hand, the single slope of the $I(q)$ for $q \parallel [100]$ is almost constant with respect to the magnetic field [Fig. 5(b)].

At 300 K, for $q \parallel [110]$, a similar field dependence is observed at the low $q$ region, where the two slopes merge at high fields, except for the presence of the bump structure at the high $q$ region [Fig. 5(c)]. The bump structure becomes clearer as the strong scattering intensity from the power-law component is suppressed by the field. For $q \parallel [100]$, the slope of $I(q)$ becomes slightly gentler under the field while the high-$q$ bump structure hardly changes [Fig. 5(d)].

The exponents $n$ obtained from the power-function fitting at 2 K and 300 K are plotted against magnetic field in Fig. 6. Here, $n$ in the low-$q$ region below $q_0$ for $q \parallel [110]$ is denoted as $n_{[110]\text{-L}}$ (red closed circles), $n$ in the high-$q$ region above $q_0$ for $q \parallel [110]$ as $n_{[110]\text{-H}}$ (blue closed squares), and $n$ for $q \parallel [100]$ as $n_{[100]}$ (green open triangles). At 2 K, $n_{[110]\text{-L}}$ gradually increases from 3.1, and $n_{[110]\text{-H}}$ gradually decreases from 5.5, both approaching 4, as the field is increased up to the saturation field around 0.5 T [Fig. 6(a)]. On the other hand, $n_{[100]}$ always stays around 4 in the whole field region up 0.5 T. This magnetic field dependence at 2 K indicates a gradual suppression of fractality of the magnetic domain structure by the field while maintaining the anisotropy.

At 300 K, $n_{[110]\text{-L}}$ increases from 3.1 to 3.4, and $n_{[110]\text{-H}}$ decreases with increasing the magnetic field to the saturation field around 0.3 T [Fig. 6(b)]. Above 0.25 T, the two power functions merge and $n_{[110]\text{-H}}$ is not well defined. In contrast to the 2 K data, $n_{[100]}$ remains around 4 at low fields but starts to decrease above 0.2 T, reaching almost the same value as $n_{[110]\text{-L}}$. These plots at 300 K clearly show a field-induced change from anisotropic strong fractal to isotropic weak fractal. This field dependence is qualitatively consistent with the MFM results, where the fractal domain pattern becomes isotropic above 0.2 T [Fig. 2(i-k)].

### 3.2.4. Broad peak structure

Finally, we analyse the high-$q$ bump structure observed only at 300 K [Figs. 5(c) and 5(d)] and consider its possible origin in Fig. 7. As shown in Fig. 7(a, b), for each of $q \parallel [110]$ and $q \parallel [100]$, the broad peak component was extracted by subtracting the power-law contribution from $I(q)$ and fitted to a Gaussian function. In both directions, the broad peak is robust against the magnetic field and observed up to 0.4 T, while its centre shifts to lower $q$ above 0.3 T, which is near the boundary between the multi-domain and single-domain states. These results suggest that the broad SANS peak is ascribed to an isotropic spatial distribution of inhomogeneously magnetized regions, whose magnitude is distinct from those in the matrix domains as schematically depicted in Fig. 7(c). Similar arguments have been reported for Fe-Ga alloys (Laver *et al*., 2010; Gou *et al*., 2021). The average spacing of the magnetic inhomogeneities, defined as $\xi = 2\pi/q_G$ (here, $q_G$ is the centre of the Gaussian peak), is plotted against field in Fig. 7(d). In both directions, $\xi \sim 40$ nm is almost independent of the field and starts to increase above 0.3 T. The robustness of the broad peak against the field can be well explained by the fact that the magnetization process at low fields is governed by the displacement of domain walls. Once the single domain state is attained and the field is further increased, some of the inhomogeneous magnetizations are forced to order and those that remain are sparse. The disappearance of the broad peak at 2 K may be accounted for by the uniform magnetization when the ferromagnetic ordering is fully developed at low temperatures. Further studies from both high-resolution real-space and reciprocal-space experiments at various temperatures are needed to characterise this disordered magnetic state in more detail.

## 4. Conclusions

We have investigated the fractal magnetic domain structure in bulk single crystals of $(Fe_{0.63}Ni_{0.30}Pd_{0.07})_3P$, a unique antiskyrmion material with $S_4$ symmetry, using SANS and MFM. The SANS intensity profile shows a power-law behaviour with an exponent close to 3, demonstrating the existence of a fractal magnetic domain structure in the bulk on the length scale down to the domain wall width where the slope of the intensity profile changes. The fractal domain structure at zero field is highly anisotropic, with the scattering intensity existing mainly along the [110] and [$\bar{1}$10] directions, and gradually disappears with increasing the magnetic field. At 2 K, the anisotropic feature remains until the fractal domain structure completely disappears at high fields, whereas, at 300 K, the anisotropic fractal structure changes to an isotropic fractal one as magnetic field is increased. Furthermore, the SANS profile shows an additional broad peak only at 300 K that is isotropic and robust against magnetic fields, suggesting inhomogeneous magnetizations within the domains. The present study provides a quantitative understanding of the anisotropic fractal magnetic domain structure with various length scales in bulk $(Fe_{0.63}Ni_{0.30}Pd_{0.07})_3P$, and demonstrates the prominent capability of SANS for studying fractal structures in magnetic systems.

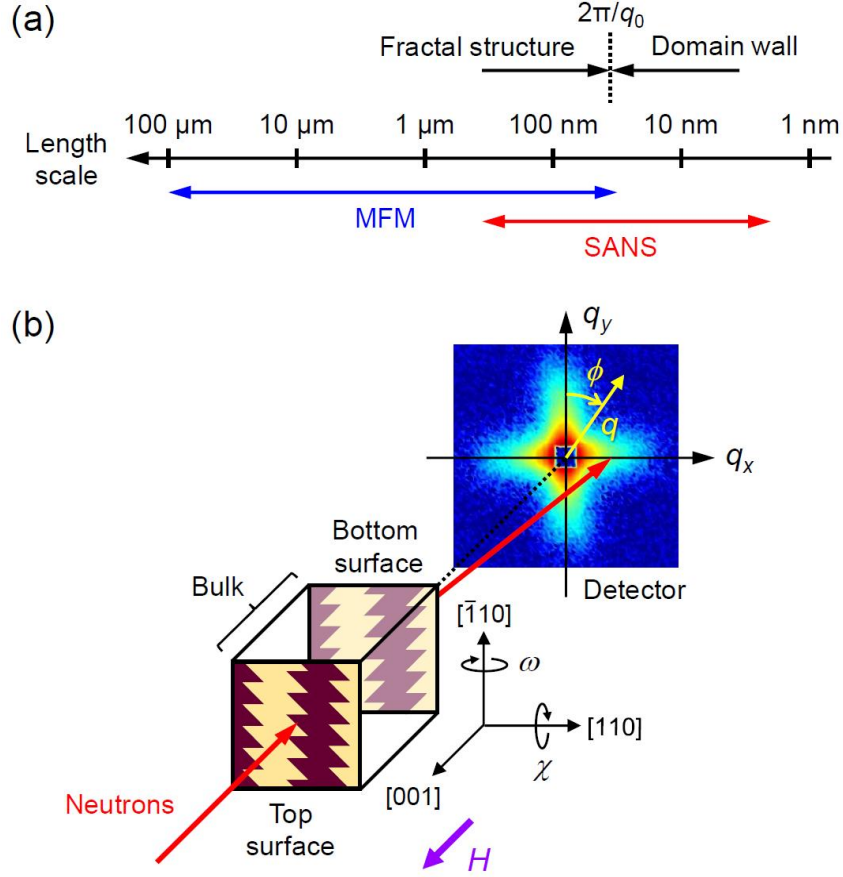

**Figure 1** (a) The length scales of fractal magnetic domain structure and domain wall, as well as the typical length scales accessible by MFM and SANS. (b) Schematic of the SANS experiment and a bulk single crystal with sawtooth magnetic domain patterns near (001) surfaces. Rocking angles $\omega$ and $\chi$, and an azimuthal angle $\phi$ are also indicated.

**Table 1** Magnetic parameters; easy magnetization axis ($T > T_{SR}$), Curie temperature $T_C$, spin-reorientation temperature $T_{SR}$, saturation magnetization $M_s$ (at room temperate), and uniaxial magnetic anisotropy constant $K_u$ (at room temperate), of $Nd_2Fe_{14}B$ (Herbst, 1991), $Mn_{1.4}PtSn$ (Vir *et al.*, 2019; Cespedes *et al.*, 2021), and $(Fe_{0.63}Ni_{0.30}Pd_{0.07})_3P$ (Karube *et al.*, 2022)

| Material | Space group | Easy axis | $T_C$ (K) | $T_{SR}$ (K) | $M_s$ (kAm$^{-1}$) | $K_u$ (kJm$^{-3}$) |
|---|---|---|---|---|---|---|
| $Nd_2Fe_{14}B$ | $P4_2/mnm$ ($D_{4h}$) | [001] | 585 | 135 | 1270 | 4600 |
| $Mn_{1.4}PtSn$ | $I\bar{4}2d$ ($D_{2d}$) | [001] | 392 | 170 | 400 | 171 |
| $(Fe_{0.63}Ni_{0.30}Pd_{0.07})_3P$ | $I\bar{4}$ ($S_4$) | [001] | 398 | None | 474 | 28 |

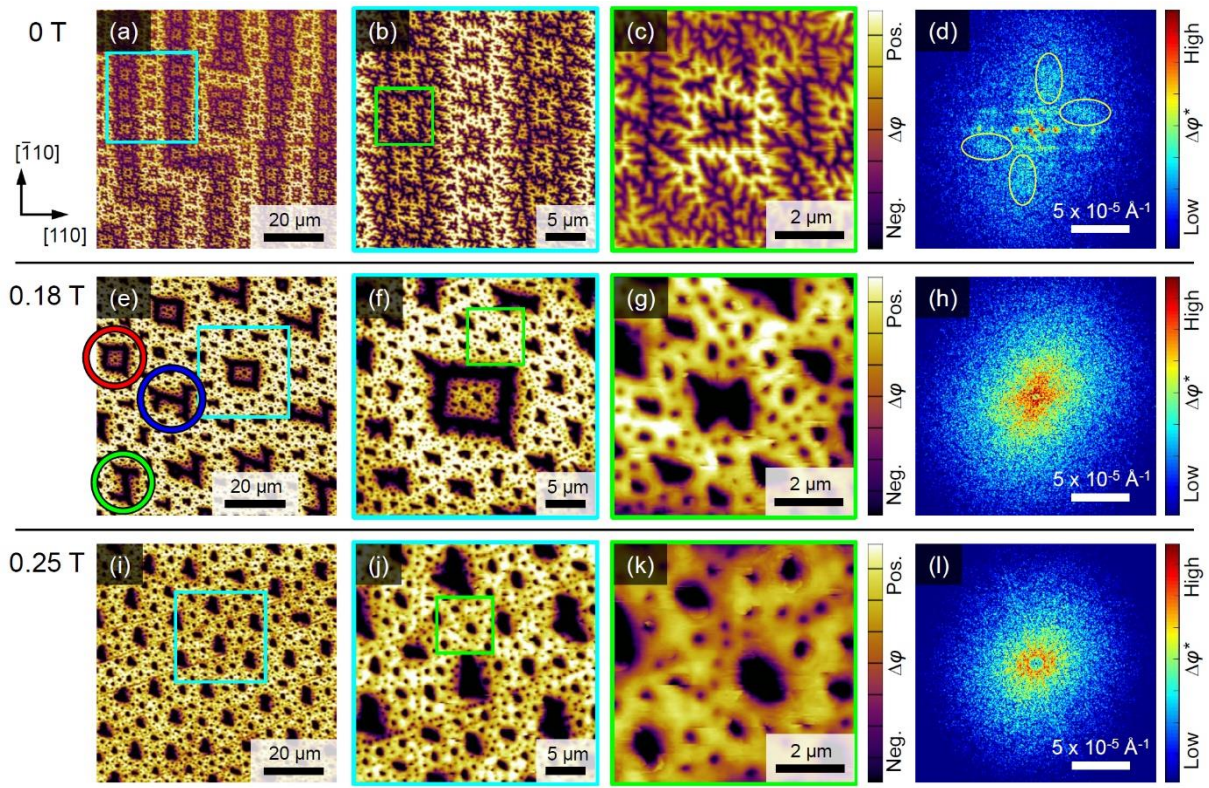

**Figure 2** MFM imaging of magnetic domain patterns at the surface of bulk $(Fe_{0.63}Ni_{0.30}Pd_{0.07})_3P$ at room temperature. MFM images at (a-c) 0 T, (e-g) 0.18 T, and (i-k) 0.25 T. Panels (b, f, j) show the higher-magnification images of the light-blue region in (a, e, i), and panels (c, g, k) correspond to the further magnified images taken for the light-green region in (b, f, j). The red, blue and green circles in panel (e) indicate square, letter N- and letter J-shaped domains, respectively. The color of the MFM images shows the phase shift of the oscillating cantilever ($\Delta\varphi$), corresponding to the stray magnetic fields perpendicular to the sample surface produced by the magnetization. Panels (d, h, l) present the corresponding FFTs of (a, e, i). The yellow circles in panel (d) highlight the asymmetric FFT intensity distribution which reflects the sawtooth pattern as governed by $\bar{4}$ symmetry.

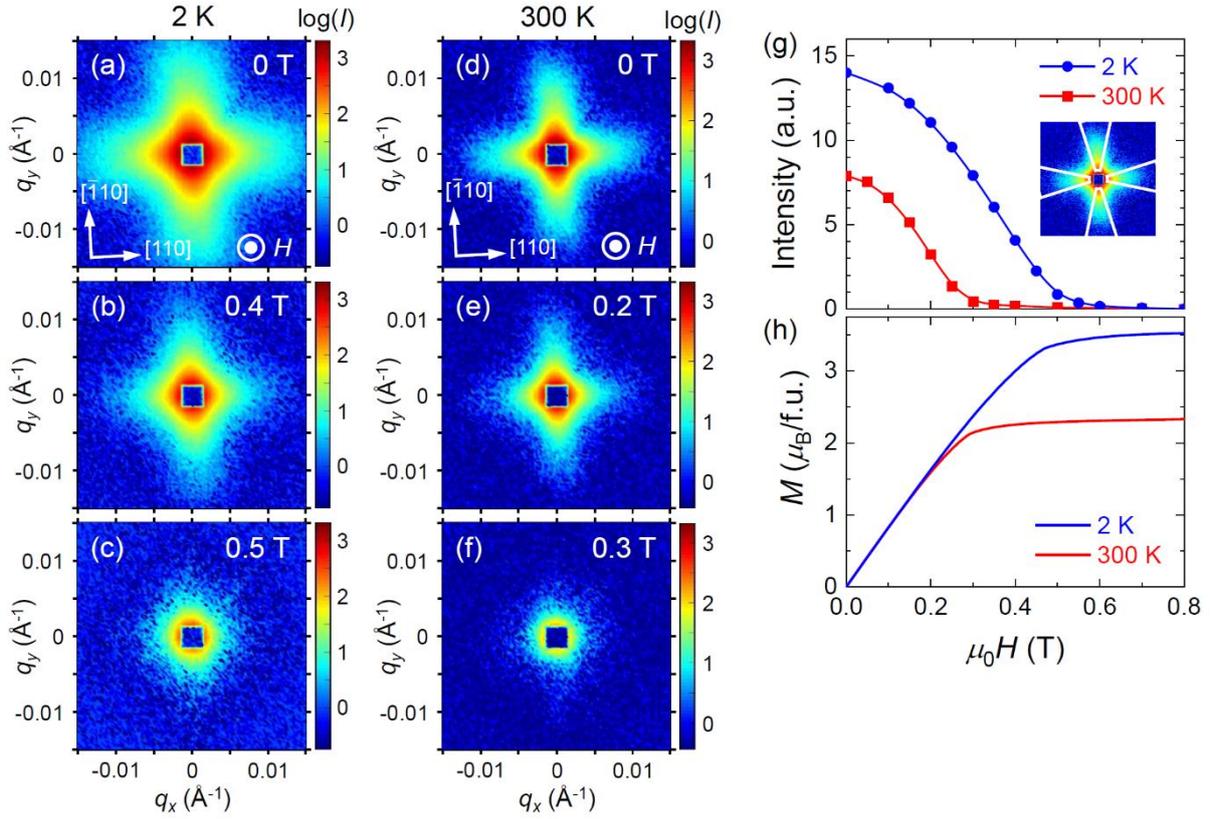

**Figure 3** SANS patterns in bulk $(Fe_{0.63}Ni_{0.30}Pd_{0.07})_3P$ at (a-c) 2 K and (d-f) 300 K at selected magnetic fields. In these SANS images, the non-magnetic scattering from the background was subtracted using the data in the fully polarized state at 0.8 T (Supporting information Fig. S3). (g, h) Magnetic field dependence of (g) the integrated SANS intensity and (h) magnetization $M$ at 2 K and 300 K. The SANS intensities are integrated over the regions along the [110] and [$\bar{1}$10] directions [white sectors in the inset of (g)].

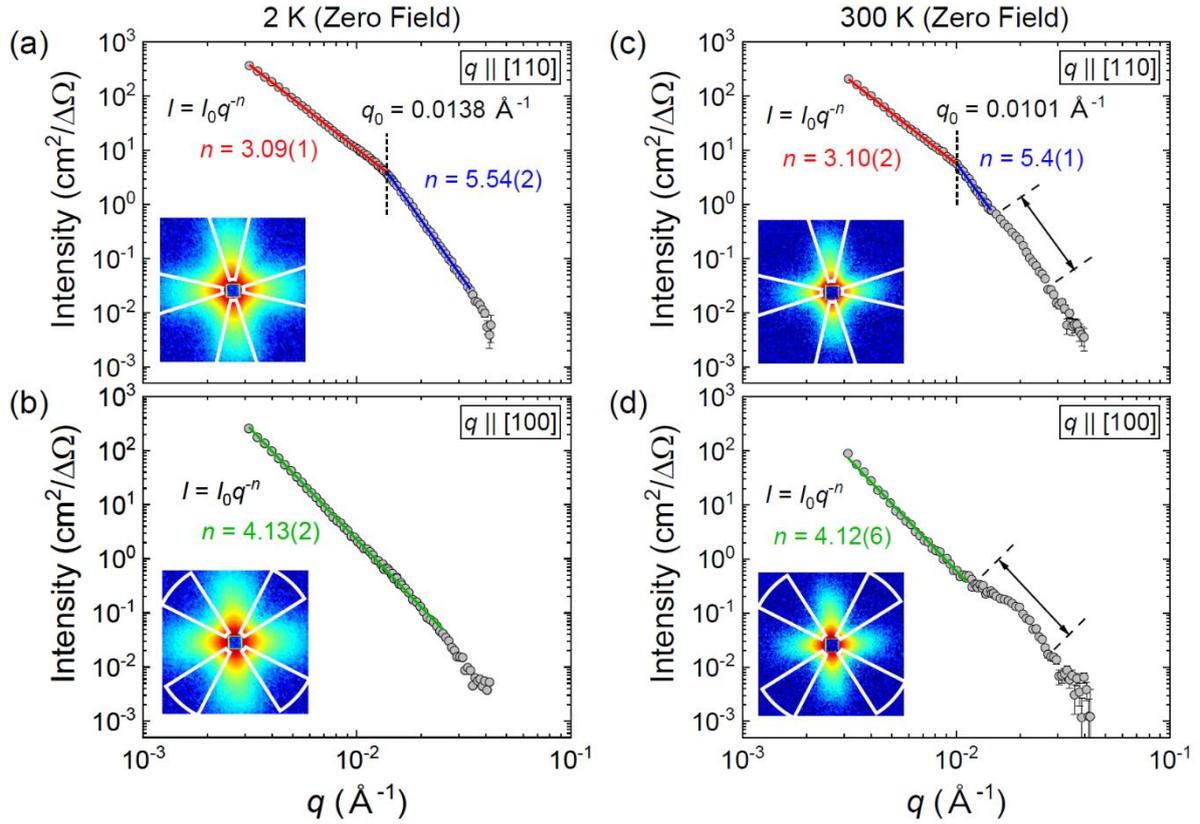

**Figure 4** Log-log plot of azimuthally averaged SANS intensity as a function of $q$ at (a, b) 2 K and (c, d) 300 K at zero field. SANS intensities are averaged over the region along (a, c) the [110] and [$\bar{1}$10] directions (azimuthal angle at $\phi = 357°, 87°, 177°, 267°$, with the width of $\Delta\phi = 30°$), and (b, d) the [100] and [010] axes ($\phi = 42°, 132°, 222°, 312°, \Delta\phi = 30°$), as indicated with the white sectors in the inset of each panel. The experimental data (grey symbols) are fitted to the power function $I = I_0 q^{-n}$ (solid lines). The crossover momentum $q_0$ where $n$ sharply changes for $q \parallel [110]$ is indicated with a vertical dashed line in panels (a) and (c). The range in $q$ of the bump structure observed at 300 K is indicated with an arrow between two dashed lines in panels (c) and (d).

**Table 2** Power exponent $n$ and crossover wavevector $q_0$ at zero field obtained from SANS measurements in $Nd_2Fe_{14}B$ (Kreyssig *et al.*, 2009), $Mn_{1.4}PtSn$ (Sukhanov *et al.*, 2020), and $(Fe_{0.63}Ni_{0.30}Pd_{0.07})_3P$ (this study).

| Material | $T$ (K) | $n$ ($q < q_0$) | $n$ ($q > q_0$) | $q_0$ (Å$^{-1}$) |
|---|---|---|---|---|
| $Nd_2Fe_{14}B$ | 200 | 3.09(5) | 4.84(6) | 0.018 |
| | 20 | 3.73(3) | - | - |
| $Mn_{1.4}PtSn$ | 250 | 2.65 | 5.8 | 0.013 |
| | 90 | 3.65 | - | - |
| $(Fe_{0.63}Ni_{0.30}Pd_{0.07})_3P$ | 300 | 3.10(2) | 5.4(1) | 0.0101 |
| | 2 | 3.09(1) | 5.54(2) | 0.0138 |

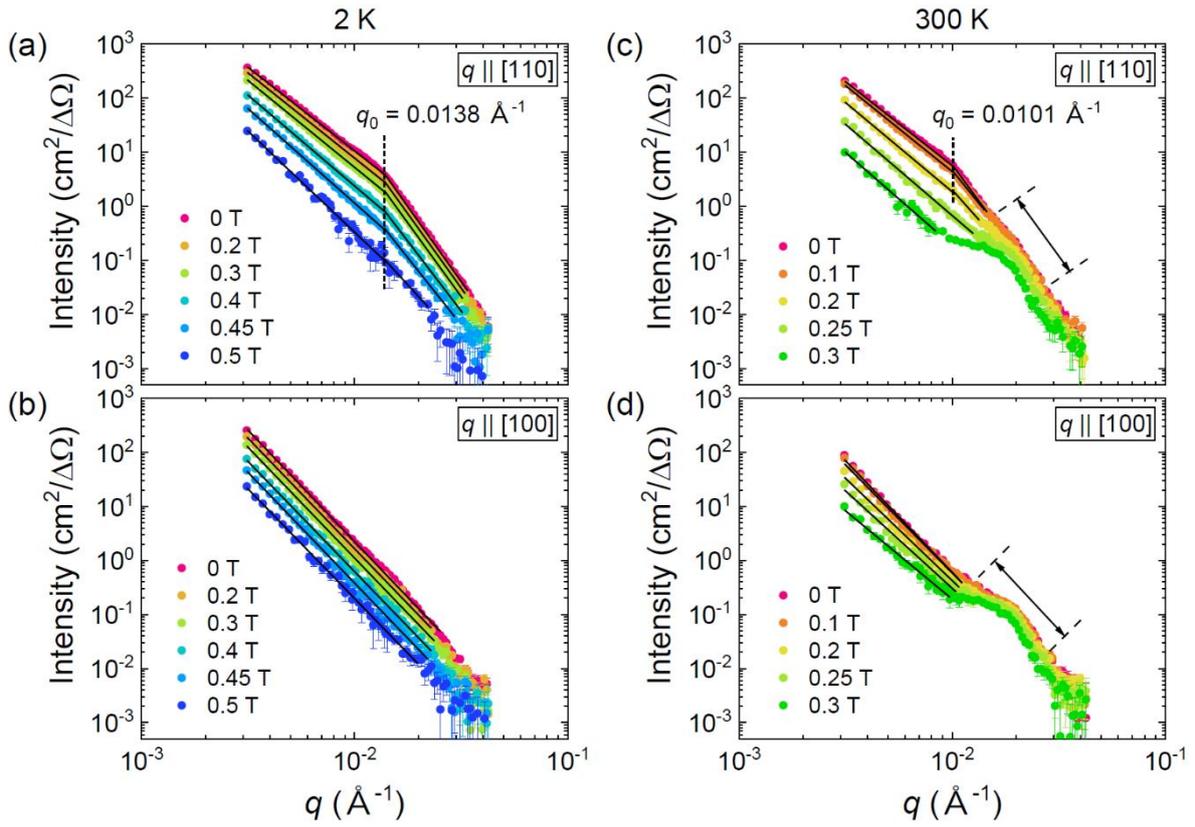

**Figure 5** Log-log plot of azimuthally averaged SANS intensity as a function of $q$ at (a, b) 2 K and (c, d) 300 K at various magnetic fields. SANS intensities are averaged over the region along (a, c) the [110] and [$\bar{1}$10] directions, and (b, d) the [100] and [010] directions, as defined in Fig. 4. The experimental data (closed symbols) are fitted to the power function (black solid lines). The crossover momentum $q_0$ for $q \parallel [110]$ is indicated with a vertical dashed line in panels (a) and (c). The range in $q$ of the bump structure observed at 300 K is indicated with an arrow between two dashed lines in panels (c) and (d).

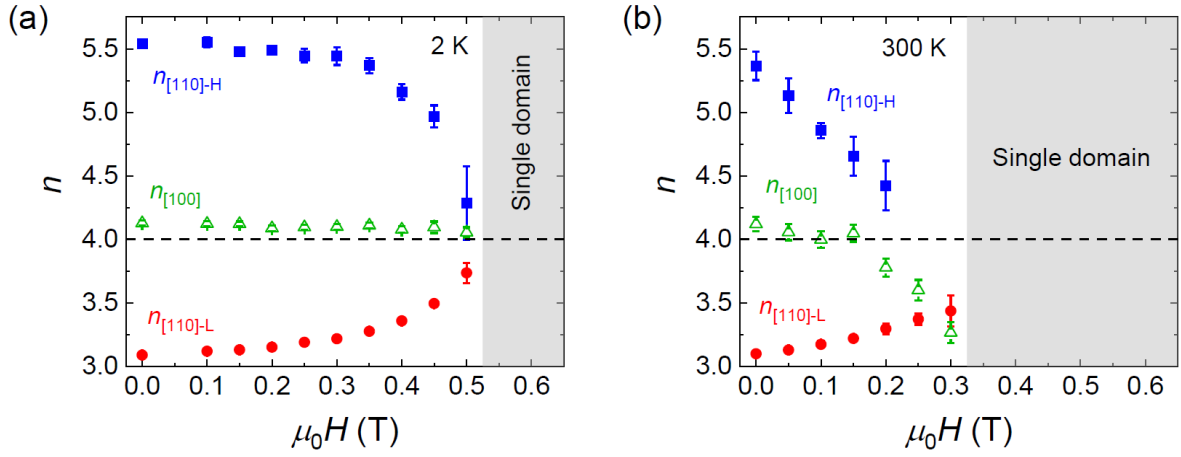

**Figure 6** Magnetic field dependence of the power-law exponent $n$ at (a) 2 K and (b) 300 K. The value of $n$ in the region of $q < q_0$ ($q > q_0$) for $q \parallel [110]$ is defined as $n_{[110]\text{-L}}$ ($n_{[110]\text{-H}}$) and shown by red (blue) closed symbols, while $n$ for $q \parallel [100]$ is defined as $n_{[100]}$ and indicated with green open symbols. The field region of the single domain state is indicated with grey shading.

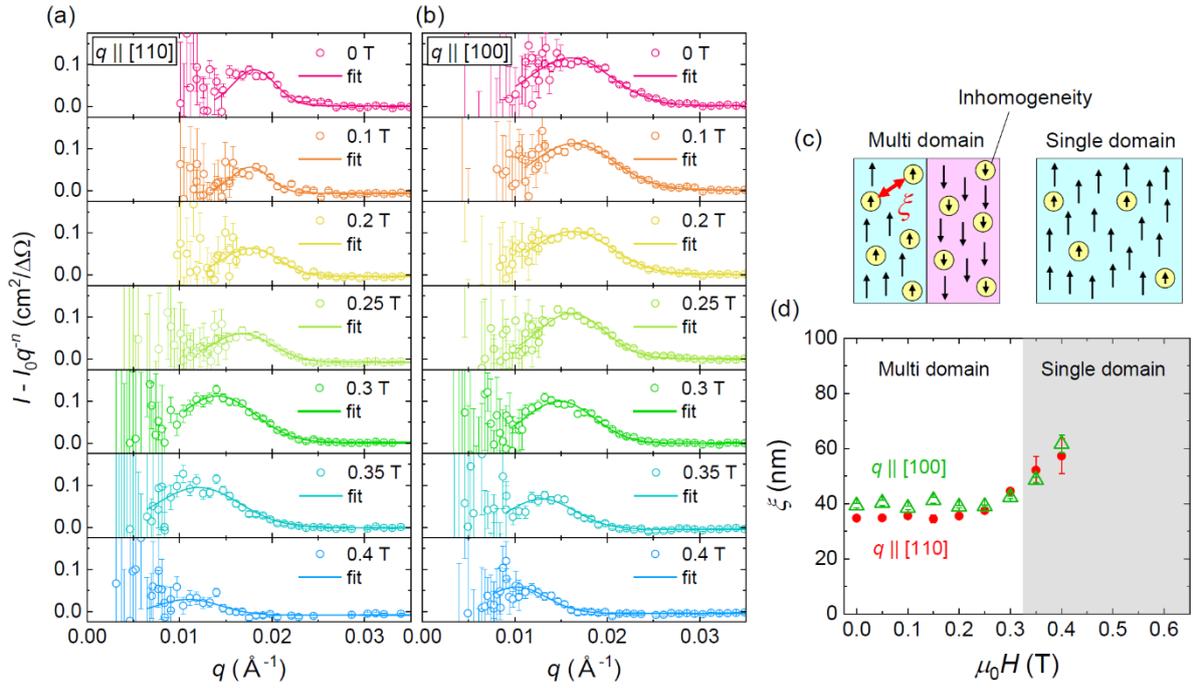

**Figure 7** Analysis of the broad SANS peak at 300 K. (a, b) Residual SANS intensity after the subtraction of the power-law component at 300 K and various magnetic fields for (a) $q \parallel [110]$ and (b) $q \parallel [100]$. The experimental data (open symbols) are fitted to a Gaussian function (solid line). (c) Schematic figures of proposed field-dependent magnetic inhomogeneities (yellow regions) within the matrix magnetic domains (light-blue and pink regions). (d) Magnetic field dependence of the average spacing $\xi$ ($= 2\pi/q_G$) of the magnetic inhomogeneities for $q \parallel [110]$ (red symbols) and $q \parallel [100]$ (green symbols), where $q_G$ is the peak centre of the Gaussian fitting. The grey shading represents the field-induced single domain region.

**Acknowledgements**   The authors thank X. Z. Yu and T. Arima for fruitful discussions. This work was supported by JSPS Grant-in-Aids for Scientific Research (Grant No. 20K15164), JST CREST (Grant No. JPMJCR20T1 and JPMJCR1874), and the Swiss National Science Foundation (SNSF) Sinergia Network CRSII5_171003 NanoSkyrmionics and Project No. 200021_188707.

# Supporting information

## S1. Experimental details of MFM measurements

In the MFM measurements, a plate-shaped bulk single crystal with a flat (001) surface was used, as shown in Fig. S1(b). Magnetic fields were applied perpendicular to the sample plate using cylindrical Nd-Fe-B permanent magnets as illustrated schematically in Fig. S1(a). Figures S1(c-j) show MFM images of fractal magnetic domain patterns at various magnetic fields. As the magnetic field is applied, the dark stripe domains shrink at 0.13 T and change to square, letter N- and letter J-shaped domains at 0.18 T. Upon further increasing the field from 0.22 T to 0.31 T, the main domains become smaller and rounder (or triangular), and all the bubble-like domains finally disappear above 0.35 T to enter a single domain state. After the magnetic field of 0.37 T is removed, many square-shaped domains remain at zero field, different from the initial stripe domains.

## S2. Box-counting analysis of MFM image

The fractal dimension of the MFM image at zero field was calculated using a box-counting method (Falconer, 1990; Smith *et al.*, 1996; Han *et al.*, 2002; Lisovskii *et al.*, 2004), as shown in Fig. S2. As this method requires an image with a sufficiently wide scale and high resolution, here a high-resolution MFM image with a size of 30 $\mu$m × 30 $\mu$m (1113 × 1113 pixels) [Fig. S2(a)] was used. For the image processing and box counting, the software ImageJ (Ver. 1.53k) was used. The MFM image was grey-scaled and binarized to extract the domain boundaries. As schematically shown in Fig. S2(b), the domain boundaries are covered with a grid of boxes of side length $s$, and the number of boxes $N$ intercepted by the domain boundaries is counted for various $s$. The fractal dimension (box-counting dimension) $D$ is defined as,

$$D = -\lim_{s \to 0} \frac{\log[N(s)]}{\log(s)}, \tag{S1}$$

and obtained from the slope of the log-log plot of $N(s)$. Figure S2(c) shows the result of the log-log plot of $N(s)$ for the MFM image. The slope at the large $s$ region is close to $D = 2$ (dotted line), but this is the dimension of the whole image plane as the selected box sizes are too large to resolve the domain boundaries, thus this segment should be excluded. A linear slope of $D = 1.29(1)$ can fit the data points in the small $s$ region between $s = 54$ nm (2 pixels) and 323 nm (12 pixels), over which fractality can be defined. Note that this length scale is comparable to that probed by SANS.

## S3. Experimental details of SANS measurements

In order to cover a wide $q$ range, SANS patterns were taken in two configurations with different collimator and detector distances, as shown in Figs. S3(a-c) and S3(d-f), respectively. Figures S3(a, d) show SANS patterns at 300 K and 0 T before background subtraction. Figures S3(b, e) show SANS patterns in the fully polarized state at 0.8 T, which show asymmetric streaks due to specular reflection of neutrons from sharp sample edges as shown in the photo [Fig. S3(g)]. The SANS patterns at 0 T after background subtraction using the data at 0.8 T are shown in Figs. S3(c, f). While the SANS signal derived from magnetic domains is strong compared to the background signal, all of the analysis presented in the main text was done on the SANS data obtained after background subtraction.

**S4. Rocking scans of SANS measurements**

Figure S4 shows the rocking scans around the vertical [$\bar{1}$10] direction (rocking angle $\omega$) and the horizontal [110] direction (rocking angle $\chi$) in the range from -3° to 3°. In both orientations, the rocking curves are very sharp (FWHM ~ 0.5°). This confirms that the cross-shaped SANS pattern is present only within the tetragonal basal plane with a long correlation length along the easy [001] direction.

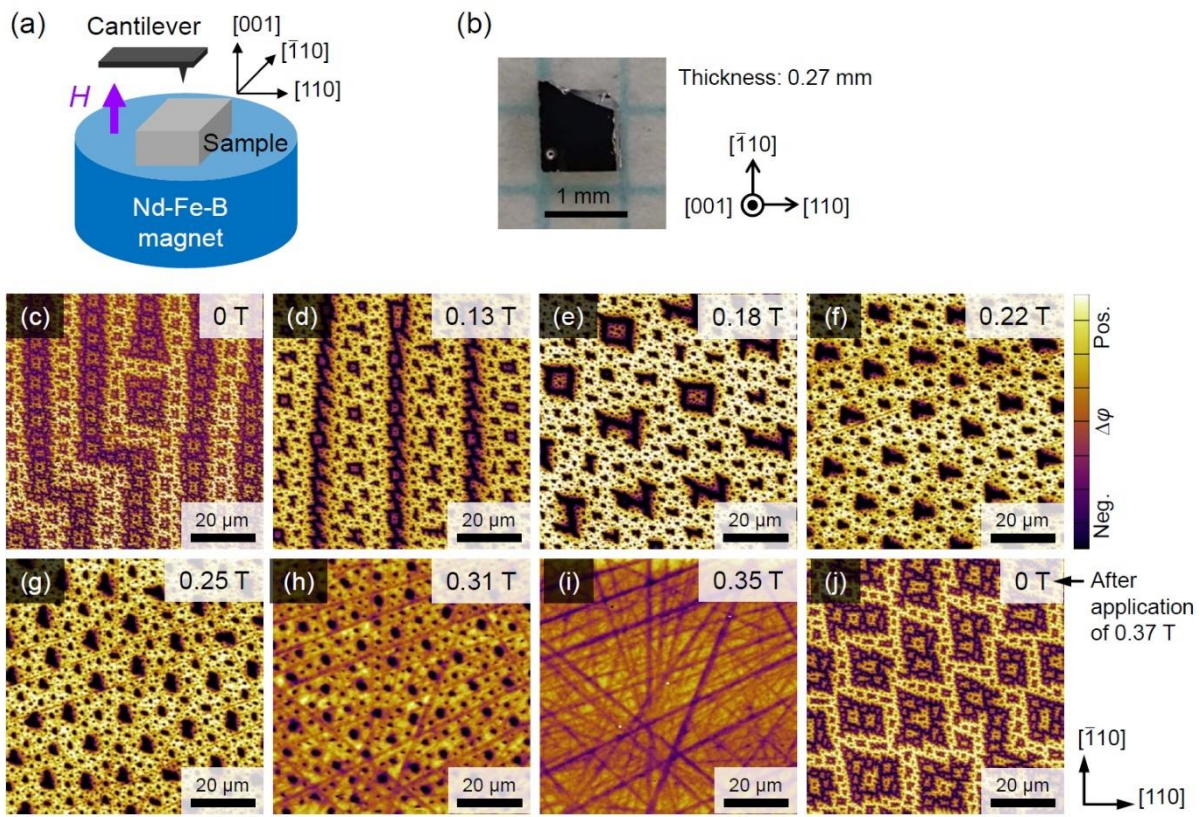

**Figure S1** Details of the MFM measurement. (a) Schematic of the experimental setup. (b) Photo of the bulk single crystal of $(Fe_{0.63}Ni_{0.30}Pd_{0.07})_3P$ used for the MFM measurement. (c-j) MFM phase images at room temperature and at various magnetic fields. A number of straight lines observed at high fields are due to polishing scratches left on the sample surface.

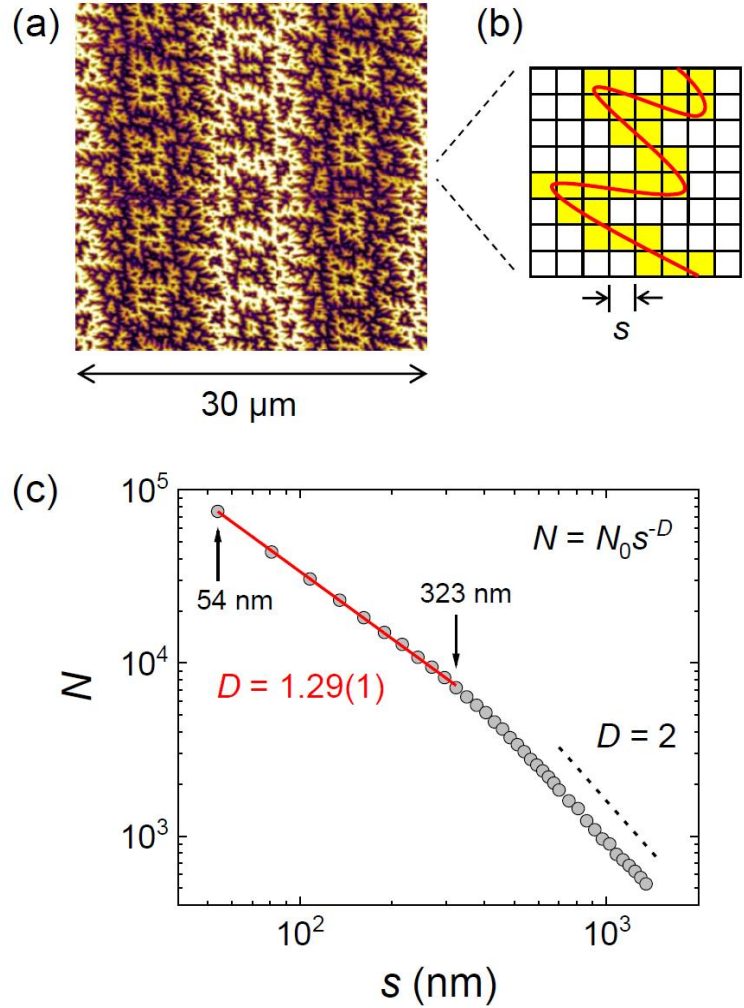

**Figure S2** Fractal analysis of the MFM magnetic domain pattern. (a) MFM image at room temperature and zero field [the same as Fig. 2(b) in the main text] used for the analysis. (b) Schematic of box-counting analysis. The number of yellow boxes $N$ intercepted by the red domain boundary is counted for various box sizes $s$. (c) Log-log plot of $N(s)$ for the MFM image. The fractal dimension $D = 1.29(1)$ is obtained from the slope (red line) over the length region between 323 nm and 54 nm as indicated with the arrows. The slope of $D = 2$ is also indicated with the dotted line.

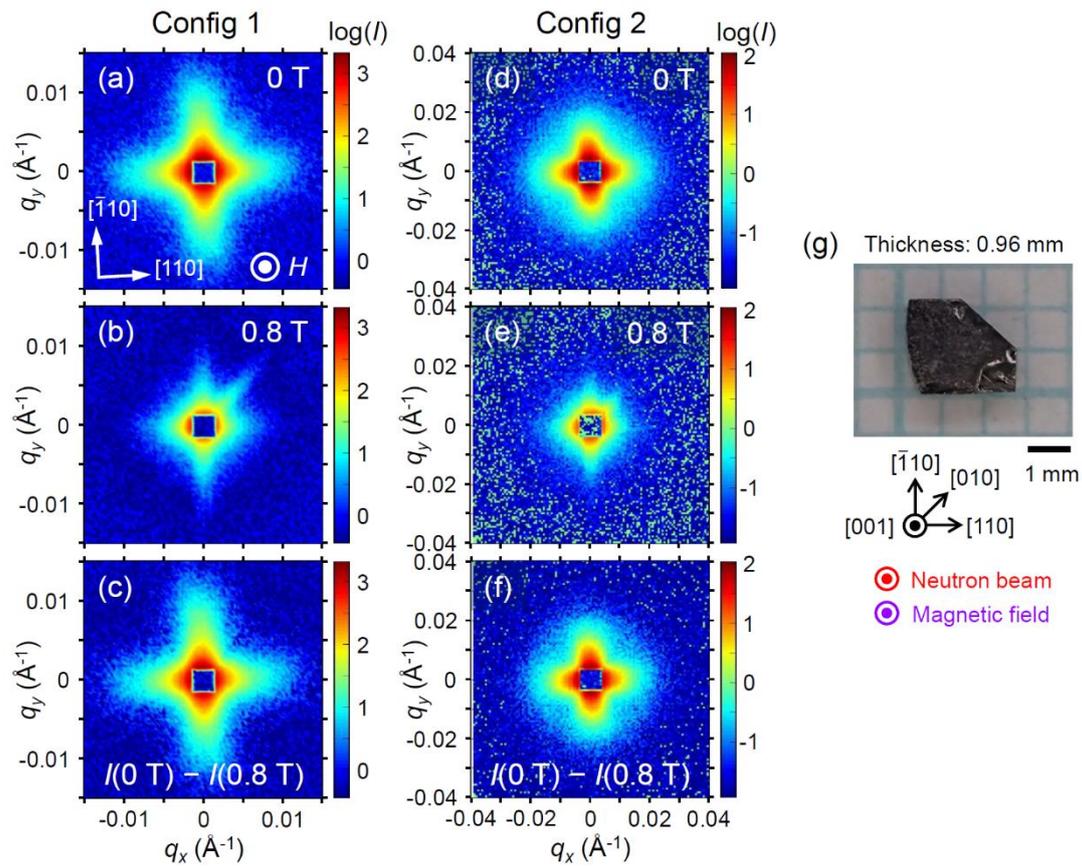

**Figure S3** Details of the SANS measurement. (a-f) SANS patterns at 300 K in two different configurations: Config 1 (detector 20 m, collimator 18 m) and Config 2 (detector 8 m, collimator 8 m). Panels (a, d) and (b, e) are SANS patterns without background subtraction recorded at 0 T (multi-domain state) and 0.8 T (fully polarized state), respectively. Panels (c, f) show the results of the subtraction of the background obtained at 0.8 T from the data at 0 T and represent the purely magnetic scattering component at 0 T. (g) Photo of the bulk single crystal of $(Fe_{0.63}Ni_{0.30}Pd_{0.07})_3P$ used for the SANS measurement.
25

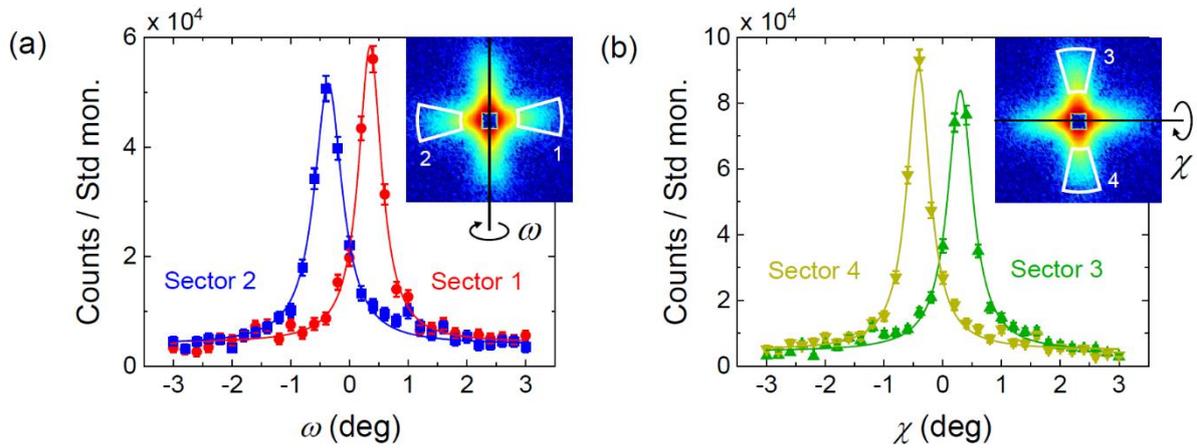

**Figure S4** Rocking scans of the magnetic SANS pattern at 300 K and zero field. (a) Rocking curves for the regions 1 and 2 indicated in the inset during the rotation around the vertical axis (rocking angle $\omega$). (b) Rocking curves for the regions 3 and 4 indicated in the inset during the rotation around the horizontal axis (rocking angle $\chi$). Data points are fitted to the Lorentzian function with a full width at half maximum (FWHM) of ~ 0.5° (solid lines).